\documentclass[%
,twocolumn%
,amssymb, amsmath,nobibnotes, aps, pre]{revtex4-2}
\usepackage{bm}%
\expandafter\ifx\csname package@font\endcsname\relax\else
 \expandafter\expandafter
 \expandafter\usepackage
 \expandafter\expandafter
 \expandafter{\csname package@font\endcsname}%
\fi

\usepackage{graphicx,subfigure}
\usepackage{amssymb,amsmath,wasysym}

\DeclareGraphicsExtensions{.pdf,.png}
\graphicspath{{./figs/}{./old_figs/}}

\usepackage[usenames,dvipsnames]{color}   % if you want to highlight some part of the paper

\usepackage[normalem]{ulem} % to bar text for instance
\newcommand*\mean[1]{\overline{#1}}

\def\be{\begin{equation}}
\def\ee{\end{equation}}
\def\bpm{\begin{pmatrix}}
\def\epm{\end{pmatrix}}

\begin{document}

\title{Optimal crawling: from mechanical to chemical actuation}

\author{P. Recho$^{1}$ and L. Truskinovsky$^2$}
\affiliation{
$^1$LIPhy, CNRS--UMR 5588, Universit\'e Grenoble Alpes, F-38000 Grenoble, France\\
$^2$PMMH, CNRS--UMR 7636, ESPCI PSL, F-75005 Paris, France
}
\email{pierre.recho@univ-grenoble-alpes.fr, lev.truskinovsky@espci.fr  }

\date{\today}%

\begin{abstract} \small 
 Taking inspiration from the crawling motion of biological cells on a substrate, we consider a  physical model of self-propulsion  where the  spatio-temporal   driving can involve both, a \emph{mechanical} actuation by  active force couples, and a \emph{chemical} actuation through controlled  mass turnover. We show that the competition and cooperation between these two modalities of active driving can drastically broaden the performance repertoire of the crawler.
  When the material turnover is slow and the mechanical driving dominates,  we find  that the highest velocity   at a given energetic cost is reached when actuation takes the form of  an active force configuration propagating as a traveling wave. As the rate of material turnover increases, and the chemical driving starts to dominate the mechanical one,   such a peristalsis-type  control progressively loses its efficacy, yielding to a standing wave type driving which  involves an  interplay between  the mechanical and chemical actuation.  Our analysis  suggests a new paradigm for   the optimal design  of crawling biomimetic  robots   where   the conventional purely mechanical driving through distributed force actuators is complemented  by a distributed  chemical  control of the material remodeling   inside the force-transmitting machinery. 

\end{abstract}

\maketitle

\section{Introduction}

In living systems the crawling mode of motility is ubiquitous and its thorough understanding on both mechanical and biochemical levels  constitutes an important fundamental challenge \cite{flaherty2007mathematical,mogilner2009mathematics,aranson2016physical,svitkina2018actin}. The parallel problem of the design of soft robots that can efficiently crawl by themselves  is an equally important engineering problem \cite{Ahn2019,Chen2020,Ze2022,Ahmed2022,Wu2023,Yao2024} which  have been mostly studied from the perspective of a purely mechanical driving in the form of distributed   actuators generating ``active'' force couples \cite{mcevoy2015materials,Zhou2015, Shen2017,agostinelli2018peristaltic,Liu2022, giraldi2020periodical, yu2020crawling,Patel2023}.
In this paper, taking inspiration from  the importance of chemical processes in the  crawling motion of biological objects, we consider the situation where in addition to a mechanical actuation, the crawler material can turnover through a process driven by an out of equilibrium  chemical reaction. 

There exists two fundamentally different paradigms to represent the overdamped crawling motion of an active object on a solid substrate. One of them, which builds on older swimming theories \cite{taylor1951analysis, lighthill1960note, lauga2009hydrodynamics}, assumes that the shape of a deformable object is dynamically actuated to harvest friction forces with its environment. The motion is supported by the breaking of the time reversal symmetry of each periodic stroke \cite{desimone2012crawling}. As initially formulated by Purcell in the context of swimming at a low Reynolds number, this necessary  condition to obtain a net motion over a stroke is known as the scallop Theorem \cite{purcell1977life,shapere1989geometry,avron2008geometric}. The optimality of such motion as a function of the actuation can be quantified by the Stokes efficiency which compares how much energy is dissipated during each stroke with  the power to move the crawler center of mass at a certain velocity against the frictional background \cite{golestanian2008analytic,golestanian2008mechanical,alouges2009optimal,alouges2017purcell}. 
Limbless animals such as millipedes, caterpillars or earthworms are good natural representatives of this   category of crawlers. 

The development of  active gel theories allowing one to model the motility of various types of biological cells \cite{julicher2007active} has paved the way for an alternative  paradigm where the crawler  is constituted of a mechanical skeleton (scaffold) which can chemically turnover through its polymerization and de-polymerization. The fundamental point here is that while the skeleton is in frictional contact with the external environment, its de-polymerized building blocks are not, in the same manner as the lower part of the belt of a rolling tank-tread transmits tractions to the ground while the upper part of the belt does not. The mechanical actuation of the skeleton by distributed contractile forces then creates a steady state  flow \cite{callan2016actin}. The spatial analogue of the scallop Theorem would then state that  to ensure self propulsion  the flow of skeleton must be asymmetric with respect to the crawler center. As a result,  the friction forces on the background  will also lose their symmetry,  leading to directional motion \cite{recho2013contraction, recho2016maximum}. However, the fact that the skeleton is being permanently advected in a particular direction requires  to ``close the stroke''.  This  means that    there must   be   sources and sinks allowing  material   renewal  and depletion, where necessary through an implicit  chemical reaction \cite{recho2015mechanics}. Again, the Stokes efficiency for  such type of motion has been introduced and the corresponding optimal regimes were identified \cite{recho2014optimality}. While the implied modeling approach  is most simply formulated in the context of an object crawling on a stiff substrate,  it was also shown to be  fully capable of  explaining various regimes of  swimming in a liquid \cite{farutin2019crawling}.

The goal of the present   paper is to  bring together these two paradigms within a single simple prototypical framework. Specifically, we consider a 1D soft elastic body with free boundaries,  where  material can be chemically driven to undergo internal mass redistribution and, in addition, can be mechanically  driven through a distributed field of active stresses.  Such system may then be actuated  both \emph{mechanically} and \emph{chemically}. It  can  be seen as representing in a very simplified way the  propulsion machinery of a cell. It would be then implied that the  active mechanical stress is  controlled   by a distribution of internally driven  molecular  motors cross-linking skeleton filaments,  while the active  mass redistribution is governed by a chemical potential representing an  internal out of equilibrium chemical reservoir of monomers that can be polymerized into filament form.
In this respect, it is worth mentioning, that the chemical pathways regulating both the cytoskeleton contractility and its turnover can now be externally controlled with light in both space and time using two different optogenetics constructions \cite{Wu2009,Valon2017}. 

One of our main results is the realization that the   interplay  between the  chemical and mechanical  modalities  of active controls crucially  depends on the relative rate of the corresponding kinetic processes. Thus,  we show that  slow turnover can hinder the chemical control while large viscosity can obstruct the work of molecular motors.  We characterize  the relative importance of  chemical versus mechanical activity by a single dimensionless parameter $\lambda$,  and study its role  in the choice of the optimal  crawling gait. More precisely, we ask how the optimal actuation strategy adjusts as $\lambda$ increases from zero to infinity and whether one can identify   transitions between different crawling gaits initiated by the variation of $\lambda$. 

In particular, we show that in order to ensure optimal efficacy,  our two modalities  of external driving, mechanical and chemical, must cooperate.   In other words,  the two controlling agents must conspire if  the goal is to achieve the best \emph{performance} at a fixed total metabolic \emph{cost}. How the corresponding mechanical and biochemical pathways are organized to  reach the necessary level of coherency is beyond the scope of this work where we    neglect  the chemo-mechanical  feedbacks between the active agents and the dynamical variables representing, for instance, the flow or the density of material. 
  
To formulate the optimal control problem we assume that  our  driving mechanisms arrive with an energetic cost. We show how  the latter  can be  specified  based on some basic thermodynamic arguments.  We then use a simple close-to-equilibrium Onsager formalism  to introduce the corresponding kinetic processes.  This allows us to specify the two time scales characterizing the mechanical and chemical drivings and formally define the parameter $\lambda$ as ratio of these time scales. 

As the most basic assumption, we  associate the performance  of the propulsion with the  average velocity of the crawler and solve in two limiting cases the mathematical optimization problem delivering the best actuation strategy at a fixed energetic cost. 
In agreement with previous investigations \cite{fang2015phase, agostinelli2018peristaltic,santhosh2022optimal}, we find that in the case of slow  material turnover, the most effective time periodic actuation strategy  is a  \emph{traveling wave}  propagating along the   body. In the opposite limit,  when turnover is fast  and both mechanical and chemical actuators are relevant, the optimal driving is represented by a \emph{standing wave}.   In the latter case,  the combined chemo-mechanical actuation and the associated cooperation between chemistry and mechanics in driving the internal dynamics,  allows  the system to reach a higher performance than in the absence of turnover at the same energetic cost. 

Using a simple ansatz linking the two limiting cases, when either mechanics or chemistry dominates,   we study the  crossover between the \emph{traveling wave} type and the \emph{standing wave} type modalities of crawling as the   parameter $\lambda$ varies continuously.  We show that the implied  crossover is  accompanied  by a   switch in the allocation of the energetic  resources   from  the purely mechanical driving  towards an optimal balance between   mechanical and chemical energy inputs.  More specifically, we show that as the rate of material turnover increases, and the chemical driving starts to dominate the mechanical one,   the  commonly accepted  peristalsis-type  control progressively loses its efficacy, yielding to a standing wave type driving which now involves a constructive  interplay between  the mechanical and chemical actuation. 

In addition to offering new insights regarding the different fundamental modalities of the functioning of living matter, our analysis  can be viewed as providing a new paradigm    for  the design  of soft robots that crawl on rigid surfaces.   The main novelty is in the  replacement of  the conventional purely mechanical driving through distributed force actuators, by a chemo-mechanical driving involving the possibility to chemically activate or dis-activate the dynamic renewal  of the internal force-transmitting scaffold.

The paper is organized as follows. In Sec.~\ref{sec:model} we derive a thermodynamically consistent   chemo-mechanical model of a driven one dimensional  crawling segment with free boundaries. In Sec.~\ref{sec:dim}, we introduce the main non-dimensional parameter.  The time periodic chemo-mechanical actuation fields that control the   dynamics are introduced in Sec.~\ref{sec:dim1}. In Sec.~\ref{sec:opt}, based on the energy balance discussed in Sec.~\ref{sec:model},  we formulate the optimization problem whose solution allows one to specify the optimal chemo-mechanical control. We then study  two limitings cases where the optimal actuation protocol can be found explicitly.  Specifically, in Sec.~\ref{s:mechanical_drive}  we assume that material turnover is slow compared to mechanical relaxation while in Sec.~\ref{s:chemical_drive} we consider  the opposite limit. Finally, we introduce in Sec.~\ref{sec:chemo_mech} an ansatz for the actuation fields that interpolates the two above limits. It enables us to study the interplay between the chemical and mechanical actuation when the relative rate of turnover varies continuously and identify a transition between the two major gaits of chemo-mechanical crawling. Our conclusions are summarized  in Sec.~\ref{sec:conclusion}.

\section{The model}\label{sec:model}

We consider a  prototypical model of a bio-mimetic object crawling along a one-dimensional track on a rigid substrate. The skeleton of this  one dimensional crawler is effectively represented as a continuum segment with material points indexed by the actual spatial coordinate \mbox{$x\in [l_-(t),l_+(t)]$}. Here the two moving boundaries representing the  front and rear  edges of the crawler are $l_-(t)$ and~$l_+(t)$ with  $t\geq 0$ denoting the time.  The deformed state of the system is described by the time dependent functions $x(X,t)$, where by $X$ we denote the reference positions of the actual points.

In view of the anticipated role of the chemically-driven mass turnover of the skeleton, we describe the mechanical response  of the system   using the framework of morpho-elasticity \cite{rodriguez1994stress, ambrosi2011perspectives}. We start by introducing  the deformation gradient $F(x,t)=\partial x/ \partial X$ and  decompose $F$ multiplicatively into  
$$F=AG.$$ 
Here $A(x,t)$ is the 1D  analog of  the elastic distortion which can be   considered in this setting as purely volumetric. We suppose that  such a distortion is counted from an unstressed configuration with a  fixed  density  $\rho_m$.  Then the actual (current) density of the material is $\rho(x,t)=\rho_m/A(x,t)$. The scalar function $G(x,t)$ describes the stress free  swelling due to the   arrival  or departure of  the material ``building blocks'' needed to  assemble  the   configuration with the density $\rho_m$. Such configuration can then be interpreted as   intermediate because the primordial   configuration of the ``building blocks'' would be  characterized by another  density  $\rho_0(x,t)=G (x,t)\rho_m $.   The function  $G(x,t) $ is  a 1D analog of the growth tensor in morpho-elasticity describing the  swelling  of the material from density $\rho_0(x,t)$ to the density $\rho_m$ while  the scalar $A$  is the deformation gradient  necessary to accommodate such a swelling   into the actual stressed configuration.  The thermodynamical configuration of the resulting    material capable of both, elastic deformation and inelastic renewal,  is described  by two mechanical (dynamic) variables which may be either $A(x,t)$ and $G(x,t)$ or $\rho(x,t)$ and $ \rho_0(x,t)$.

\emph{ Mass balance.}  In the Eulerian coordinate system, the equation of  mass  balance  can be rewritten  in the form
\begin{equation}\label{e:mass_bal_0}
\partial_t\rho+\partial_x(\rho v)  = r,
\end{equation} 
where 
$$d/dt=\partial_t+v\partial_x,$$  $v=\partial_tx(X,t)$ is the velocity of material points,  and the mass supply is 
$$r=\rho \left(\frac{1}{G} \frac {d G}{dt} \right),$$
see e.g.   \cite{goriely2017mathematics}. It is clear from  \eqref{e:mass_bal_0} that it is the \emph{temporal} variation of $G$ which brings  the local sources/sinks of mass. For instance, in the context of living cell, this term represents the polymerization and depolymerization of the   cytoskeleton filaments which is controlled by a monomers reservoir  \cite{pollard2022cell}. As we do not consider any flux of mass through the   boundaries, we shall also assume that $$\dot{l}_{\pm}=v(l_{\pm}(t),t).$$

\emph{Momentum balance. }  Since inertia is negligible in our setting,  we can write force balance in the form 
\begin{equation}\label{e:force_bal_0}
\partial_x\sigma=f_f,
\end{equation}
where  $\sigma(x,t)$ is the axial  stress and  $f_f$ is the bulk force describing  the  interaction of the skeleton with the rigid substrate in a thin film limit. We further assume that 
\begin{equation}\label{e:force_bal_01}
 \sigma(x,t)= \sigma_e(x,t)+\sigma_a(x,t),
\end{equation}
where $\sigma_e(x,t)$ is the elastic stress and $\sigma_a(x,t)$ is an active stress, representing the mechanical driving and effectively describing  the momentum  exchange with an out of equilibrium   reservoir. One can also think directly in terms of the driving bulk force and the elastic restoring force:
$$f_a(x,t)=\partial_x\sigma_a \text{ and } f_e=-\partial_x\sigma_e.$$ 
The former  can  be exogenous  (say, originating from  external actuators such as an applied magnetic field affecting embedded beads) or endogenous (say, describing  myosin molecular motors cross-linking actin filaments).
The boundary condition associated with  \eqref{e:force_bal_0} is  
$$\sigma(l_{-}(t),t)=\sigma(l_{+}(t),t),$$  which implies the existence of  a  stiff spring  representing a constraint which   connects the two edges of the crawling system $l_{+}, l_{-}$ and ensures that  its total length 
$$L=l_{+}(t)-l_{-}(t)$$ 
remains constant. 

\emph{Energy balance.} We consider that the system is \emph{isothermal} and introduce  its free energy  in the form
\begin{equation} \label{energy}
\Psi=\int_{l_-}^{l_+}\rho \psi(A,G) \mathrm{d}x,
\end{equation}
where we assume that elasticity and turnover are  uncoupled material properties of the skeleton: $$\psi(A,G)=\psi_e(A)+\psi_c(G).$$ For simplicity, we consider as variable only the mechanical contribution $\psi_e(A)$ while the chemical contribution  
 $\psi_c(G)=\tilde{\psi}$ is assumed constant. This reflects the fact that only variations of the elastic strain change the stored internal energy while the amount of stress-free swelling due to material turnover does not impact the chemical free energy as it is often formulated in classical morpho-elastic theories \cite{erlich2023mechanical}. In view of the local mass conservation \eqref{e:mass_bal_0} we  obtain:
\begin{align*}
\frac{d\Psi}{dt}=\int_{l_-}^{l_+}\left(r\psi+\rho\frac{d\psi}{dt}\right) \mathrm{d}x=\int_{l_-}^{l_+}(\sigma_e\partial_xv+r\mu_c)\mathrm{d}x,
\end{align*}
where we introduced the elastic stress in the skeleton
$$\sigma_e=\rho_m\frac{\partial\psi_e}{\partial A},$$ 
and the  chemical  potential of the skeleton
$$\mu_c=\psi-A\frac{\partial\psi_e}{\partial A}.$$

Next we need to introduce  the exerted power $\Pi$. Since we assumed for simplicity  that there are no boundary traction forces (no cargo to carry), such a  power is delivered only in the bulk, through both active stresses and  active mass exchange.
The power of active forces  acting on the system   has the standard form of mechanical work per unit time 
$$\Pi^m=\int_{l_-}^{l_+}f_av\mathrm{d}x.$$
Similarly, the power of generalized forces performing regulation of mass exchange and acting at the chemical level,   can be written as 
$$\Pi^c=\int_{l_-}^{l_+}\mu_a r \mathrm{d}x. $$
The external mass reservoir is characterized by the  chemical potential $\mu_a(x,t)$ which is a direct chemical analog of $f_a(x,t)$, while the rate of the exchange reaction $r(x,t)$ is the analog of the local velocity $v(x,t)$.

Due to the fact that our system is active, the integral  energy inequality representing the second law of thermodynamics can be interpreted in two equivalent forms.  

First, considering $f_a$ and $\mu_a$ as internal drivings, we can write the integral  energy  balance as  $-d\Psi/dt=\hat R$ where   the function $\hat R=R-(\Pi^m+\Pi^c)$ describes all energy exchanges due to the interaction between  the  material system described by the energy \eqref{energy} and the machinery operating  to maintain the active terms.   The latter   includes the  rate of supply of  mechanical and chemical energy due to microscopic active  agents $\Pi^m+\Pi^c$ and the irreversible dissipation $R \geq 0$ characterizing the interaction of the system with a thermal reservoir.   Note that if the active agents are designed to always operate in the regime  where $\Pi^m+\Pi^c\geq 0 $, as will be the case in the rest of the paper, the corresponding entry  in $\hat R$ can be considered as representing  ``anti-dissipation''. 
 
 The second approach, which we effectively use in what follows, is based on the assumption that  the power  $\Pi^m+\Pi^c$ represents an external  work. We can then   write the integral  energy balance in the form 
\begin{align}\label{e:dissipation}
R&=\Pi^m+\Pi^c-\frac{d\Psi}{dt}\\
&=\int_{l_-}^{l_+}\left[(f_a-f_e)v+ \left( \mu_a-\mu_c\right) r \right].\nonumber
\end{align}
Here the dissipation $$R\geq 0,$$  is interpreted as the difference between the work done per unit time   by the macroscopic active agents $\Pi^m+\Pi^c$ and the concurrent rate of  change of the macroscopic free energy of the system $d\Psi/dt$,  see e.g.   \cite{recho2014optimality, deshpande2021chemo}.  

To ensure that the term $R$ representing  the dissipated energy, is non-negative as required by the second law of thermodynamics,   we make the simplest assumption  that  the linear Onsager close-to-equilibrium theory \cite{de2013non} is operative. Specifically, we assume that the thermodynamic fluxes and forces are linearly related so that 
\begin{equation}\label{e:dissipation11} 
\begin{array}{lcl} 
f_a-f_e=\xi v\\
\mu_a-\mu_c=\nu r.
\end{array}
\end{equation}
In \eqref{e:dissipation11},  we have neglected chemo-mechanical cross terms and introduced the purely  mechanical friction coefficient $\xi>0$ and the purely chemical kinetic coefficient  $\nu>0$.   Under these assumptions,   friction with the  substrate, mimicking interaction with the mechanical reservoir, and kinetic turnover of the  material,  mimicking interaction with the  chemical reservoir, represent two  sources of quadratic dissipation in the system. As we show below, the two   parameters, $\xi$ and  $\nu$,  set the \emph{timescales} of the corresponding mechanical and chemical relaxation processes and their dimensionless ratio is an important control parameter of the problem.

\emph{Specification of the model. } We  further impose for simplicity that the free energy  represents only entropic elasticity and set 
\begin{equation}\label{e:dissipation12}
\psi_e(A)=\psi_0(A\log(A)-A),
\end{equation}
where $\psi_0$ characterizes the elastic stiffness of the material. The chosen expression \eqref{e:dissipation12}  can be viewed as reflecting a skeleton containing almost ideal polymer chains. Note that  $\partial_A\psi(1)=0$ which agrees with our  assumption that after the mass exchange the reference configuration with density $\rho_m$  remains  stress-free. Under the assumption \eqref{e:dissipation12}, the elastic stress-strain relation takes the form
\begin{equation}\label{e:const_0}
\sigma_e=-E\log\left( \frac{\rho}{\rho_m}\right) 
\end{equation}
where $E=\rho_m\psi_0$ is the  Young modulus at the stress-free state. Expression \eqref{e:const_0} penalizes both infinite polymer chains compression and extension.  The resulting  equation $\eqref{e:dissipation11}_1$ can be written as
\begin{equation}\label{e:const_01}
f_a-E\frac{\partial_x\rho}{\rho}=\xi v. 
\end{equation}
Analogously we can now write an explicit expression  for the thermodynamic chemical potential
\begin{equation}\label{e:const_02}
\mu_c=\tilde{\psi}+ \frac{E}{\rho},
\end{equation}
which allows us to rewrite the resulting  equation $\eqref{e:dissipation11}_2$ in the form
\begin{equation}\label{e:const_03}
 \mu_a-\tilde{\psi}+ \frac{E}{\rho} =\nu r.
\end{equation}
Equation \eqref{e:const_01} effectively describes the kinetics of the mechanical relaxation of the internal flow $v$ while equation \eqref{e:const_03} describes the kinetics of the chemical relaxation of the material turnover;   the two equations are coupled through the mass balance equation \eqref{e:mass_bal_0}.  

\section{Non-dimensionalization}\label{sec:dim}

To analyze the resulting system of equations, it is convenient to introduce  the change of variable:
$$y=\frac{x-S(t)}{L}\in [-\frac{1}{2},\frac{1}{2}] \text { where, }  S(t)=\frac{l_-(t)+l_+(t)}{2}$$
is the geometric center of the segment. We use $L$ to non-dimensionalize space, $\xi L^2/E$ to non-dimensionalize time, $\rho_m$ to non-dimensionalize density and $E/L$ to non-dimensionalize force. Combining \eqref{e:mass_bal_0},\eqref{e:const_01} and \eqref{e:const_03}, the problem reduces to a single dimensionless reaction-drift-diffusion equation for the skeleton density field $\rho(x,t)$:
\begin{equation}\label{e:model_nd}
\partial_t\rho+\partial_y\left(\rho (f-V)-\partial_{y}\rho\right) =\frac{1}{\tau}\left(\frac{1}{\rho}-\frac{1}{\tilde{\rho}}+\mu \right), 
\end{equation}
with the boundary conditions
\begin{equation}\label{e:model_bc_nd}
\rho\vert_{-1/2}=\rho\vert_{1/2} \text{ and } V(t)=\left( f-\frac{\partial_y\rho}{\rho}\right)\vert_{\pm 1/2}.
\end{equation}
In \eqref{e:model_nd}-\eqref{e:model_bc_nd}, $$V=\dot{l}_{\pm}$$ 
is the macroscopic velocity of the crawler,  while  $f=f_a/(E/L)$ and   $\mu=\mu_a/(E/\rho_m)$ are dimensionless active spatio-temporal controls.
 
We also introduced two non-dimensional  parameters. The first one  
\begin{equation}\label{e:const_04}
\tau=\frac{\nu \rho_m^2 }{\xi L^2}, 
\end{equation}
represents the ratio of the two characteristic time scales: the one characterizing the  chemical reaction controlling the skeleton turnover and  the other one  describing  mechanical relaxation due to the sliding friction against the rigid background. The second dimensionless parameter compares the stored elastic energy compared to the chemical stored energy
\begin{equation}\label{e:const_05}
\tilde{\rho}=\frac{E}{\rho_m\tilde{\psi}}.
\end{equation}
In Sec.~\ref{s:weak_actuation}, we show that as we shall be interested in small material density variations only,  the  only important dimensionless parameter, controlling the choice of the optimal crawling strategy will be the combination of the two parameters \eqref{e:const_04} and \eqref{e:const_05}: 
\begin{equation}\label{e:const_06}
\lambda=\frac{1}{\tau \tilde{\rho}^2}=\frac{\xi\tilde{\psi}^2 L^2}{\nu E^2 }.
\end{equation}
 Thus when $\lambda=0$, material turnover is absent and sliding friction with the substrate is the only source of dissipation and therefore the main  rate limiting process.  Instead,  when $\lambda\gg 1$, turnover is much faster than  frictional relaxation and the energy entering the system is predominantly dissipating due to the chemical reaction.

\section{Actuation}\label{sec:dim1}

 The free boundary problem \eqref{e:model_nd}, with the boundary conditions \eqref{e:model_bc_nd}, also contains two dimensionless active fields, which are still not specified.  One is the active force distribution $f(y,t)$ and  the other one is the active chemical potential  distribution $\mu(y,t)$. Both of them represent  non-equilibrium reservoirs driving the system, a mechanical one and a chemical one, respectively.
 
 Given that intrinsic mechanical action  is exerted by force couples, it can be convenient to write 
$$f=-\partial_y s,$$ 
where  the potential $s(y,t)$ is the  active stress  \cite{kruse2005generic}. In the context of ATP-driven acto-myosin systems, it describes  the field of force dipoles  generating mechanical contraction. Instead, the  field $\mu(y,t)$ represents the dynamic target of an out-of-equilibrium chemical reaction  responsible for the internal turnover of matter.   

Observe that when both fields vanish and therefore, the driving is absent, the solution of \eqref{e:model_nd}-\eqref{e:model_bc_nd}  is: $\rho\equiv\tilde{\rho}$ and $V \equiv 0$. Two non-equilibrium limiting cases are of interest:    $\mu=0$,  corresponding to a  purely mechanical driving  and $s=0$ when  the actuation is purely  chemical.

One possibility to set $s(y,t)$ and $\mu(y,t)$ is to impose that they are controlled by the presence of chemo-mechanical feedbacks so that the driving ``follows'' the responding system as the two are described by a coupled system of equations. For instance, in some models of   contraction-driven cell motility, the actuation process is coupled with the dynamics of molecular motors which, in turn,   is  linked to  the flow of cytoskeleton \cite{recho2014optimality}.
Other types of coupling involving a feedback relating material flow with the active stress describing various regulatory pathways have been considered, for instance in \cite{nishikawa2017controlling, qin2018biochemical, blanchard2018pulsatile}. Physically motivated systems involving the coupling between active turnover and an intra-cellular chemical messenger affected by the material flow have been considered as well \cite{giverso2018mechanical}.
 %If we further assume that   both,  the variations of  density $\rho$ compared to the equilibrium density $\tilde{\rho}$ are small and the temporal fluctuations of actin compared to the turnover dynamic  are  small as well (see \cite{putelat2018mechanical} for details), then     \eqref{e:model_nd} simplifies giving  $\rho-\tilde{\rho}=-\tau\tilde{\rho}^3\partial_yv$.

\section{Optimal driving}\label{sec:opt}

In this paper we are not assuming any of the aforementioned feedbacks  and instead ask the question of how to reverse engineer the crawling machinery to reach the optimal performance. One can think of our system as describing a soft crawling body driven either by   internal (cell motility) or external (robotics) agents.  The task is to understand how an  \emph{optimal performance} can be achieved when it can be  driven using both mechanical -$f(y,t)$- and chemical -$\mu(y,t)$- actuation. We leave the question, whether such controls can be indeed \emph{implemented} in a self-consistent  and physically meaningful manner,  to a separate study.

Some works have previously addressed  the issue of cell motility in the framework of  control theory  by resorting to chemo-mechanical controls in specific situations \cite{carlsson2011mechanisms, Recho2015} but without considering the energetic cost of such actuation. The present paper complements and further develops this previous work.  

 The first step is to restrict the class of admissible controls $f(y,t)$ and $\mu(y,t)$.   To reflect a cyclic nature of the actuation, we assume   that these controls  are  $T$-periodic in time where $T$ is a positive number to be found in the process of finding the solution of the optimization problem. Since we do not take into account  any resultant part   in either  mechanical and chemical drive, it is natural to assume that the time and space averages of the controls  $f(y,t)$ and $\mu(y,t)$ are  equal to zero:
\begin{equation}\label{e:expansions1}
 \langle f\rangle=\langle \mu\rangle= \bar{f}=\bar{\mu}=0. 
\end{equation}
Here 
$$\left\langle (.) \right\rangle=\lim_{T\rightarrow \infty}\frac{1}{T}\int_0^{T}(.)\mathrm{d}t \text { and } \mean{(.)}=\int_{-1/2}^{1/2}(.)\mathrm{d}y,$$ 
denote the time averaging and the space averaging respectively. 

First recall  that the total time averaged power injected into the system as a result of both mechanical and chemical activity   is 
\begin{equation}\label{e:expansions11}
 C=\overline{\left \langle \Pi^m+\Pi^c \right\rangle}=\overline{\left \langle fv+\mu r \right\rangle}.
 \end{equation}
By averaging \eqref{e:dissipation} over time, we obtain that $C\geq 0$, showing that the total actuation cost, expressed in this way,  is always non-negative. This also shows that the power of the active driving is indeed anti-dissipative over each stroke.  Substituting the expressions for $v$ and $r$ from the Onsager constitutive relations \eqref{e:const_01}-\eqref{e:const_03} into \eqref{e:expansions11}, we can rewrite the expression for the measure of the \emph{energetic cost}  in the form 
$$C =  \left\langle \overline{f\left(f-\frac{\partial_y\rho}{\rho}\right)}\right\rangle +  \frac{1}{\tau}\left\langle \overline{\mu\left(\mu+\frac{1}{\rho}-\frac{1}{\tilde{\rho}}\right)}\right\rangle. $$ 
The definition of the \emph{functional performance} is less straightforward since the system can move even in the absence of cargo.  Naturally, in this case one would   like to associate  a non-zero functionality  even if the  resultant applied force is equal to zero. To circumvent this classical problem, several proposals have been made. Among them  we consider the Stokes performance  \cite{Lighthill1952}.  In our case  this means  choosing  the  rate of frictional dissipation $\xi L \left\langle V\right\rangle^2$ necessary to advance  the system as a rigid object  as  the functionality  measure. When both the length $L$ of the system  and the viscosity  coefficient $\xi$ are fixed, this choice is equivalent to associating the performance of the crawling mechanism with the averaged velocity 
$$P=\left\langle V\right\rangle.$$
 Other choices are possible as well, accounting  for instance, for the metabolic expenses required to maintain at a certain level the distributed  active stresses, see for instance the discussion in \cite{garcia2019guided}.
 
In this paper we have chosen to  set the problem of maximizing  the performance $P$ at a fixed energetic cost $C$.  This optimal control problem, also considered by \cite{ agostinelli2018peristaltic,santhosh2022optimal}, reflects the desire to compare  crawler  designs with an imposed availability of metabolic resources and select the design that achieves the best performance. Other choices would be possible such as minimizing the cost at a fixed performance, which would correspond to selecting, among  the crawlers that achieve a given performance, the design that consumes the less resources. It is also possible to maximize the efficiency $P/C$ (see \cite{recho2014optimality}) which corresponds to an optimal trade-off between the cost and the performance. The choice between these and others seemingly arbitrary options is ultimately dictated by the targeted functionality of the crawler.

 %$C=\left\langle ( f^*)^2\right\rangle +\tau\left\langle (\partial_t \rho^*)^2\right\rangle\geq 0$ 

\section{Weak actuation}\label{s:weak_actuation}

To get analytic results, we only consider the case where the spatial and temporal  inhomogeneities of  the driving forces are small. Then if  $\epsilon$ is a small parameter characterizing the scale  of activation, we can write 
\begin{align}\label{e:expansions}
&s(y,t)= \epsilon  s_1(y,t)+\epsilon^2  s_2(y,t)+...\\
&\mu(y,t)= \epsilon  \mu_1(y,t)+\epsilon^2  \mu_2(y,t)+...\nonumber
\end{align}
To assess the cost $C$ and the performance $P$,  we shall need to  compute the first two  terms in both  expansions. To this end,  it will be convenient to introduce the new auxiliary variables:
 $$ u=\mu \tilde{\rho},$$ characterizing chemical control and $$w=s+\log\left(   \rho/\tilde{\rho}\right),$$ mixing the mechanical control with mechanical response. We can then also write the associated expansions:
 \begin{align}\label{e:expansions_2}
&u(y,t)=\epsilon  u_1(y,t)+\epsilon^2  u_2(y,t)+...\\
&w(y,t)=\epsilon  w_1(y,t)+\epsilon^2  w_2(y,t)+...\nonumber
\end{align}
Inserting \eqref{e:expansions_2} in \eqref{e:model_nd}-\eqref{e:model_bc_nd}, the first order problem  takes the form
\begin{equation}\label{e:first_order}
\partial_tw_1-\partial_{yy}w_1+\lambda w_1=\partial_ts_1+\lambda(s_1+u_1)
\end{equation}
with the periodic boundary conditions
\begin{equation}\label{e:first_order_bc}
w_1\vert_{-1/2}=w_1\vert_{1/2} \text{ and } V_1(t)=-\partial_yw_1\vert_{\pm 1/2}.
\end{equation}
In the context of weak actuation, our choice of the auxiliary variables $u$ and $w$ leads to the appearance in \eqref{e:first_order} of only one non dimensional parameter $\lambda$ introduced in \eqref{e:const_06}.
Since $\left\langle u_1\right\rangle=\left\langle s_1\right\rangle=0$, the  first non vanishing contribution to  the cost is of second order
 $C=\epsilon^2C_2+...$ where 
  \begin{equation}\label{e:perf1}
  C_2=\left\langle\overline{\partial_yw_1\partial_ys_1}\right\rangle+ \lambda \left\langle\overline{ u_1 (s_1+u_1-w_1)}\right\rangle.
  \end{equation}
As it involves only first order terms, $C_2$ can be computed from the solution of \eqref{e:first_order}-\eqref{e:first_order_bc}.

Similarly, since $\left\langle V_1\right\rangle=0$, the first non vanishing contribution to the  performance is also second order $ P=\epsilon^2 P_2+...$ where
 \begin{equation}\label{e:perf}
P_2=\left\langle V_2\right\rangle.
\end{equation}

To compute this contribution we need to consider the second order expansion  for  $w_2$:
\begin{align}\label{e:second_order}
&\partial_tw_2-\partial_{yy}w_2+\lambda w_2=\partial_ts_2+\lambda(s_2+u_2)+\nonumber\\
&(\partial_yw_1+V_1)(\partial_yw_1-\partial_ys_1)+\nonumber\\
&\lambda(s_1-w_1) (2u_1-3w_1+3 s_1)/2,
\end{align}
with the periodic boundary conditions
\begin{equation*}\label{e:second_order_bc}
w_2\vert_{-1/2}=w_2\vert_{1/2} \text{ and } V_2(t)=-\partial_yw_2\vert_{\pm 1/2}.
\end{equation*}
Then, after performing the temporal averaging of these equations over time we obtain
\begin{align}\label{e:second_order_aver}
-\partial_{yy}\left\langle w_2\right\rangle+\lambda \left\langle w_2\right\rangle &=\left\langle(\partial_yw_1+V_1)(\partial_yw_1-\partial_ys_1)\right\rangle\\
&+\frac{\lambda}{2}\left\langle(s_1-w_1) (2u_1-3w_1+3 s_1)\right\rangle,\nonumber
\end{align}
with $P_2$ entering the boundary conditions:
\begin{equation}\label{e:perf_2}
\left\langle w_2\right\rangle\vert_{-1/2}=\left\langle w_2\right\rangle\vert_{1/2} \text{ and } P_2=-\partial_y\left\langle w_2\right\rangle\vert_{\pm 1/2}.
\end{equation}
Again, as the right hand side of \eqref{e:second_order_aver} only involves first order terms, it is sufficient to consider the first order contributions of the controls $s_1$ and $u_1$ in order to obtain both $C_2$ and $P_2$.

Our problem is to   maximize the functional $P_2$ at  a given  value of the functional $C_2$. To identify  admissible solutions of this problem  we also need to impose that 
\begin{equation}\label{e:ineq_const}
\left\langle \overline{f_1^2+u_1^2}\right\rangle \leq 1 
\end{equation}
which  insures that the  spatio-temporal variations of the obtained solutions are in agreement with the  asymptotic  expansion \eqref{e:expansions}. 
The upper bound in \eqref{e:ineq_const}, which we chose to be equal to one  controls the magnitude of the active fields $f_1$ and $u_1$, up to a renormalization of $\epsilon$.

Note that since we maximize $P_2$, we focus exclusively on positive values of the velocity $V_2$. In view of the spatial symmetry of the system \eqref{e:first_order}-\eqref{e:first_order_bc}, in addition to  an actuation protocol ($u_1(y)$, $s_1(y)$) that  gives the performance $P_2$, there is always an  actuation protocol ($u_1(-y)$, $s_1(-y)$) delivering the  performance $-P_2$ at the same cost.  So the solution for the  negative velocity corresponding to minimization of $P_2$ can be obtained from our results.

As we have already mentioned, our main goal  is to investigate the role of the parameter $\lambda$, characterizing the relative importance of the chemical versus the mechanical activity, in the choice of the optimal crawling gait. More precisely, we ask how the optimal actuation strategy adjusts as $\lambda$ increases from zero to infinity.

\section{Purely mechanical driving}\label{s:mechanical_drive}

Our starting point is the purely mechanical (elastic) limit where the mass exchange with the chemical reservoir  maintaining the turnover is absent. In other words, we assume that $\lambda \to 0$ and study the limit when the reaction regulating  material turnover is much slower  than the mechanical flow of matter. In this approximation   the chemical driving $u_1$ becomes  irrelevant.

We shall use this limiting case as a benchmark for the rest of the analysis: we will not consider any energetic cost constraint in this section and only maximize the performance regardless of the cost. The obtained maximal performance will then be associated as a benchmark to a certain cost value which will be maintained constant when we study the other cases where $\lambda>0$. In this way, all the actuation protocols considered in this paper will be comparable as they will be characterized by the same level of injected power.

Using the convenient variables  $j_{1}=-\partial_y w_{1 }$ and $f_{1}=-\partial_y s_{1 }$ we can rewrite  \eqref{e:first_order}-\eqref{e:first_order_bc} in the form
\begin{equation}\label{e:flux_lin_tau_lim}
 \begin{array}{c}
\partial_{t}j_1-\partial_{yy}j_1=\partial_{t}f_1 \\
V_1=j_1\vert_{\pm 1/2} \text{ and }\partial_yj_1\vert_{-1/2}=\partial_yj_1\vert_{1/2}.
\end{array}  
\end{equation}
The corresponding expressions for the cost \eqref{e:perf1} and the performance \eqref{e:perf}
simplify accordingly:  
\begin{equation}\label{perf2}
 C_2^0=\left\langle\overline{j_1f_1}\right\rangle, \,\,\, P_2^0=\left\langle\overline{\rho_1 j_1} \right\rangle, 
 \end{equation}
where $\partial_y\rho_1=f_1-j_1$ and $\overline{\rho_1}=0$.  To justify the expression for $P_2^0$ in \eqref{perf2},  we introduce  the variable  $j_{2}=-\partial_y w_{2 }$ and  write 
$$\partial_y\left\langle j_2\right\rangle=-\left\langle\partial_y\rho_1(j_1-V_1)\right\rangle=-\left\langle\partial_y(\rho_1(j_1-V_1))\right\rangle,$$
where we used that, in view of  \eqref{e:flux_lin_tau_lim},   $\partial_t\rho_1=-\partial_yj_1$. Hence, given that  $P_2^0=\left\langle j_2\right\rangle\vert_{\pm 1/2}$, the expression in \eqref{perf2} follows.

To solve the linear heat equation  \eqref{e:flux_lin_tau_lim} we use a standard approach and represent the mechanical driving $f_1$ in  Fourier series
\begin{equation}\label{e:spatial_f1_modes}
f_1(y,t)=\sum_{l=1}^{\infty}f_1^{2l}(t)v_{2l}(y)+f_1^{2l-1}(t)v_{2l-1}(y),
\end{equation}
where we separated  the terms containing  spatially even and spatially odd modes,  
$v_{2l}(y)=\sqrt{2}\cos(2l\pi y)$ and $v_{2l-1}(y)=\sqrt{2}\sin(2l\pi y)$, respectively.
In view of the time periodic nature of the driving,  the time dependent coefficients $f_1^{2l,2l-1}$  are $T$-periodic. We can also write a similar representation for the solution of \eqref{e:flux_lin_tau_lim}
$$j_1(y,t)=\sum_{l=1}^{\infty}j_1^{2l}(t)v_{2l}(y)+j_1^{2l-1}(t)v_{2l-1}(y),$$
 whose time dependent coefficients satisfy the  equation 
$$\partial_{t}j_1^{2l,2l-1}(t)+\alpha_lj_1^{2l,2l-1}(t)=\partial_{t}f_1^{2l,2l-1}(t),$$
where $\alpha_l=4\pi^2l^2$.

 At large times, the  solution of the above ordinary differential  equation takes the form 
$$j_1^{2l,2l-1}(t)=f_1^{2l,2l-1}(t)-\int_{0}^{\infty}K_l(t-u)f_1^{2l,2l-1}(u)\mathrm{d}u,$$
where the kernel is  $K_l(t)=\alpha_{l}\exp(-\alpha_lt)\text{H}(t)$ while   $H(t)$  is the standard Heaviside function. In terms of physical variables it means that 
$$\rho_1^{2l,2l-1}(t)=\frac{-(-1)^{2l,2l-1}}{2l\pi}\int_{0}^{\infty}K_l(t-u)f_1^{2l-1,2l}(u)\mathrm{d}u.$$
We can now compute the spatial average of interest  
\begin{align*}
&\overline{\rho_1j_1}=\sum_{l=1}^{\infty}\frac{1}{2l\pi}\int_{0}^{\infty}\int_{0}^{\infty}\\
&
G_{l}(t,u,v)\left[f_1^{2l}(u)f_1^{2l-1}(v)-f_1^{2l-1}(u)f_1^{2l}(v) \right]\mathrm{d}u\mathrm{d}v.
\end{align*}
where, $G_{l}(t,u,v)=K_l(t-u)\left[\delta(t-v)-K_l(t-v))\right]$. Finally, performing the time averaging of $G_{l}$, we   obtain,
\begin{equation} \label{P2}
P_2^0=\lim_{M\rightarrow \infty}\sum_{l=1}^{\infty}
\int_{0}^{M}\int_{0}^{M}
Q_{l}(u-v)\frac{f_1^{2l}(u)f_1^{2l-1}(v)}{Ml\pi}\mathrm{d}u\mathrm{d}v,
\end{equation}
where $Q_l(t)=-\alpha_l\text{sign}(t)\exp(-\alpha_l |t|)/2$. Using the same approach we can also  compute the cost
\begin{align} \label{C2}
& C_2^0=\lim_{M\rightarrow \infty}\sum_{l=1}^{\infty}\left[ 
\int_{0}^{M}(f_{2l}(u)^2+f_{2l-1}(u)^2 )\mathrm{d}u-\right. \nonumber \\
 &
\left.\int_{0}^{M}\int_{0}^{M}L_{l}(u-v)(f_{2l}(u)f_{2l}(v)+f_{2l-1}(u)f_{2l-1}(v))\mathrm{d}u\mathrm{d}v\right] 
\end{align}
where  $L_l(t)=\alpha_l\exp(-\alpha_l |t|)/2$.

The next step is to  express  the time dependent coefficients  $f_1^{2l,2l-1}$ in temporal Fourier series:
\begin{equation}\label{e:modes_modes}
f_1^{2l,2l-1}(t)=
 \sum_{k=1}^{\infty}A_{2l,2l-1}^kv_{2k}\left(\frac{t}{T}\right)+B_{2l,2l-1}^kv_{2k-1}\left(\frac{t}{T}\right).
\end{equation}
Substituting these  expressions into \eqref{P2} and \eqref{C2} and performing integration we finally obtain
\begin{equation}\label{e:perf_tau_inf_gen}
P_2^0=\sum_{l=1}^{\infty}\sum_{k=1}^{\infty}\frac{2kT \alpha_l(A_{2l}^kB_{2l-1}^k-A_{2l-1}^kB_{2l}^k)}{l(\alpha_k+\alpha_l^2 T^2)}
\end{equation}
and
\begin{equation}\label{e:cost_tau_inf_gen}
C_2^0=\sum_{l=1}^{\infty}\sum_{k=1}^{\infty}\frac{\alpha_k[ (A_{2l}^k)^2+(A_{2l-1}^k)^2+(B_{2l}^k)^2+(B_{2l-1}^k)^2] }{\alpha_k+\alpha_l^2T^2}.
\end{equation} 
 The constraint \eqref{e:ineq_const}  reduces to 
\begin{equation}\label{e:L2_tau_inf_gen}
\sum_{l=1}^{\infty}\sum_{k=1}^{\infty}(A_{2l}^k)^2+(A_{2l-1}^k)^2+(B_{2l}^k)^2+(B_{2l-1}^k)^2\leq 1.
\end{equation}
Note that as a result of these manipulations, our original PDE control problem has been reduced to an algebraic optimization problem.

Before   moving  to the solution of this algebraic problem,   we observe  that 
for  actuation with  time reversal symmetry $f_1(t)=f_1(-t)$ all coefficients  $B^k=0$  and therefore  we have from  \eqref{e:perf_tau_inf_gen} that  $P_2^0=0$. This observation can be viewed as a variant of the Scallop Theorem \cite{purcell1977life}  in our overdamped system: the  time reversal symmetry of the actuation must  be broken for self propulsion to become possible.
 
We now  focus on  the maximization of the performance $P_2^0$ subjected to the inequality constraint \eqref{e:L2_tau_inf_gen}. 
To this end we introduce the matrix  
$$ \mathbb{Q}=\left( \begin{array}{cccc}
0&0&0&1/2\\
0&0&-1/2&0\\
0&-1/2&0&0\\
1/2&0&0&0
\end{array}\right)$$
and the vector 
$$U_{k,l}=\left( \begin{array}{c}
A_{2l}^k\\
A_{2l-1}^k\\
B_{2l}^k\\
B_{2l-1}^k
\end{array}\right). $$
 In these notations, the problem is to maximize   at each $k$ and $l$  the quadratic form $$ U_{k,l}^T\mathbb{Q}U_{k,l}=A_{2l}^kB_{2l-1}^k-A_{2l-1}^kB_{2l}^k,$$ 
under the constraint that 
\begin{equation} \label{ineq}
\sum_{l=1}^{\infty}\sum_{k=1}^{\infty} m_{k,l}\leq 1,
\end{equation}
 where $m_{k,l}=U_{k,l}^TU_{k,l}$.
The eigenspace of $\mathbb{Q}$ corresponding to its largest eigenvalue  is two-dimensional. With the inequality constraint \eqref{ineq} taken into account, this eigenspace, containing the optimal actuation modes, can be parametrized by the two sets of coefficients $a_{k,l}$ and $m_{k,l}$ and written in the form  
$$U_{k,l}=\frac{\sqrt{m_{k,l}}}{\sqrt{2}}\left( \begin{array}{c}
\cos(a_{k,l})\\
-\sin(a_{k,l})\\
\sin(a_{k,l})\\
\cos(a_{k,l})
\end{array}\right).$$ 
 Substituting  
 this expression of $U_{k,l}$ into \eqref{e:perf_tau_inf_gen}, we obtain
$$P_2^0=\sum_{l=1}^{\infty}\sum_{k=1}^{\infty}\frac{klm_{k,l}T}{k^2+\alpha_l (l T)^2},$$
which does not involve the coefficients $a_{k,l}$. The remaining problem of maximizing $P_2^0$ in  $k,l$ can be viewed as a problem of  allocating the weights $m_{k,l}$ at integer points $(k,l)$ where the expression $kl/(k^2+\alpha_l (l T)^2)$ reaches its largest value. There is actually only a single point ($k_0,l_0$) where such an expression reaches its  maximal value and it is then natural to set $m_{k_0,l_0}=1$. 
Finding the maximum of the function $(k,l)\rightarrow kl/(k^2+\alpha_l (l T)^2)$ when $k$ and $l$ are considered to be continuous variables and setting $\gamma=T/(2\pi)$, we find
$$k_0=\left\lceil \frac{1}{2} \left(\sqrt{4 \gamma^2+1}-1\right)\right\rceil$$
and 
$$l_0=\left\lceil \frac{\sqrt{3} \sqrt{\gamma \left(\gamma+2 \sqrt{\gamma^2+12}\right)}-3 \gamma}{6 \gamma}\right\rceil,$$ 
where $\left\lceil.\right\rceil$ is the ceiling function. This choice delivers the maximum of $P_2^0$ at a given $T$.  

Finally,  maximizing $P_2^0$ with respect to  this remaining parameter,  we  find that  $T=k/(2\pi)$ where $k\geq 1$ is an arbitrary integer. This  means that all such actuation protocols (with $k/T$ fixed) give  the same optimal performance value $P_2^0=1/(4\pi)$ while  the  corresponding   spatial mode is always the same one with $l=1$.   The optimal  actuation protocol is then 
\begin{equation}\label{e:TW_actuation_no_turnover}
f_1(y,t)=\sqrt{2}\cos(2\pi(2\pi t-y+a)),
\end{equation}
where $a$ is an arbitrary  phase.  Using  \eqref{e:cost_tau_inf_gen}, we find that such  performance level is reached at the cost $C_2^0=1/2$.

The resulting  optimal actuation can be characterized as a \emph{traveling wave} propagating from the rear to the front. This     actuation strategy  is in fact often observed in the motion of limbless crawlers. It has already been analytically shown to be optimal in a similar framework in \cite{ agostinelli2018peristaltic,santhosh2022optimal}. In these previous works,   which also adopted a one-dimensional setting, the definition of the cost and performance are the same as ours and the authors also consider the maximization of the performance at a fixed imposed cost. However, the friction law  is more general, allowing for a strain-dependent dissipation (which would involve the  dependence of our parameter $\xi$ on the dynamic variable $\rho$). An   important difference with  \cite{agostinelli2018peristaltic}, is that there, the active  control is imposed on a discretized reference \emph{strain} rather than our control of active \emph{stress}. In this sense the approach of  \cite{santhosh2022optimal}, also adopting that actuation is performed by internal forces, is closer to ours.  The  results in  \cite{santhosh2022optimal} can be also considered as more general
 since no \emph{a priori}  assumption of time periodicity of the mechanical actuation  is made.
 %and they study the influence of the value of the imposed cost on the maximal performance. 

Instead of focusing on these already investigated issues, in the rest of the paper  we move away from the purely mechanical problem and engage  chemical activation while  fixing the cost  at  $C_2=C_2^0=1/2$.  Thus,  we restrict  the available power of  actuation based on the  cost which  emerged from the maximal performance reachable in the purely mechanical problem.  This will allow us  to compare chemical and mechanical  activation  strategies at the same level of power delivery.  

\section{Primarily chemical driving}\label{s:chemical_drive}
We now turn  to the opposite limit  $\lambda\rightarrow \infty$ when the reaction regulating  material turnover is much faster than the mechanical flow of matter. In this limit, the reaction source term in \eqref{e:model_nd} dominates the mechanical drift-diffusion. However, mechanics is still playing a role by affecting  the  chemical potential $\mu_c$ through its strain dependence which  makes both $s_1$ and $u_1$ operative in the optimization problem. 

In order to avoid  the formation of  boundary layers, we further assume that both, the mechanical  driving  $s_1$  and the chemical driving $u_1$, satisfy  periodic boundary conditions. In this case, we obtain from \eqref{e:first_order} that $w_1-s_1+u_1\rightarrow 0$, physically corresponding to the equilibrium condition: $\mu_c=\mu_a$ (in dimensional form).

Under these assumptions we can write,  
\begin{align}\label{e:C_2_inf}
C_2^{\infty}&=\left\langle\overline{\partial_yw_1\partial_ys_1}\right\rangle+\lambda \left\langle\overline{ u_1 (s_1+u_1-w_1)}\right\rangle\nonumber\\
&=\left\langle\overline{\partial_y(u_1+s_1)\partial_ys_1}\right\rangle+ \left\langle\overline{ u_1 (\partial_tw_1-\partial_{yy}w_1-\partial_ts_1)}\right\rangle\nonumber\\
&=\left\langle\overline{\partial_y(u_1+s_1)^2}\right\rangle.
\end{align}
For the performance, \eqref{e:second_order_aver} reduces to $\left\langle r_2\right\rangle \rightarrow\left\langle u_1^2\right\rangle/2$, hence  
\begin{equation}\label{e:P_2_inf}
P_2^{\infty}=-\frac{1}{2}\partial_y\left\langle  u_1^2 \right\rangle\vert_{\pm 1/2}.
\end{equation}
The goal now is to  maximize $P_2^{\infty}$ with the constraint $C_2^{\infty}=1/2$ while respecting  the small perturbation inequality  \eqref{e:ineq_const}. 

The problem is explicitly expressed in terms of the controls without the use of the auxiliary function $w_1$.  Note also that  time  is playing a transparent role in this problem since it is involved only in the final averaging operation. It is therefore possible to first solve the optimization problem with controls that are only space dependent and then multiply the obtained solutions by any $T$-periodic function whose square average is equal to one (in order to fulfill the constraints).  In other words,   time and space variables can be separated in the optimal actuation protocols. 

We thus expand $u_1$ and $s_1$  in Fourier series using  the $v_{2l,2l-1}$ basis  
\begin{align}\label{e:spatial_u1_s1_modes}
u_1(y)&=\sum_{l=1}^{\infty}u_1^{2l}v_{2l}(y)+u_1^{2l-1}v_{2l-1}(y),\\
s_1(y)&=\sum_{l=1}^{\infty}s_1^{2l}v_{2l}(y)+s_1^{2l-1}v_{2l-1}(y),\nonumber
\end{align}
 which gives 
\begin{align}\label{e:p_inf_c_inf}
P_2^{\infty}&=-4\pi\sum_{l=1}^{\infty}l \left\langle u_1^{2l}u_1^{2l-1}\right\rangle \\
 C_2^{\infty}&=4\pi^2\sum_{l=1}^{\infty}l^2 \langle(u_1^{2l}+s_1^{2l})^2 +(u_1^{2l-1}+s_1^{2l-1})^2\rangle.\nonumber
\end{align}
The inequality constraint \eqref{e:ineq_const} takes the form
\begin{equation}\label{e:cons_inf}
\sum_{l=1}^{\infty}4\pi^2 l^2 \langle(s_1^{2l})^2+(s_1^{2l-1})^2\rangle + \langle(u_1^{2l})^2+(u_1^{2l-1})^2\rangle\leq 1.
\end{equation}
In the above formulation, all the spatial modes can be multiplied by a time dependent function $\phi(t)$, leaving the problem unchanged as soon as $\langle\phi^2 \rangle=1$.

 Following the same  approach as  in  section~\ref{s:mechanical_drive}, we first maximize the term $u_1^{2l}u_1^{2l-1}$  under the   constraints: $$4\pi^2l^2 [(u_1^{2l}+s_1^{2l})^2 +(u_1^{2l-1}+s_1^{2l-1})^2]=c_l$$ and $$4\pi^2l^2[(s_1^{2l})^2+(s_1^{2l-1})^2] + (u_1^{2l})^2+(u_1^{2l-1})^2=m_l,$$ where $c_l$ is a non-negative sequence of numbers that sums to $1/2$ and $m_l$ is a  similar sequence whose sum is smaller than $1$.
 
 The resulting mathematical problem can be again qualified as  the maximization of a quadratic form under two quadratic constraints. Maximizing the associated Lagrangian function with two Lagrange multipliers corresponding to the two constraints, we  obtain that at given $c_l$ and $m_l$, the maximal performance is  
\begin{align}\label{e:P2cons}
P_2^{\infty}(l,c_l,m_l)&=\frac{2 \pi  l}{\left(4 \pi ^2 l^2+1\right)^2} [c_l (4 \pi ^2 l^2-1)+m_l+\\
&4 \pi  l (\sqrt{c_l \left(-c_l+4 \pi ^2 l^2 m_l+m_l\right)}+\pi  l m_l)].\nonumber
\end{align}
As $P_2^{\infty}$ is decreasing as a function of $l$ (with other variables fixed) and increasing as a function of both $c_l$ and $m_l$ (also with other variables fixed), the expression \eqref{e:P2cons} is made maximal when the weights  $c_1=1/2$ and $m_1=1$ are allocated to the first spatial mode at $l=1$. 

We then obtain that the optimal performance is  
\begin{equation}
P_2^{\infty}=\frac{1}{ 1/\pi +12 \pi -4 \sqrt{1+8 \pi ^2}}
\end{equation}
and the corresponding optimal  actuation protocols are:
\begin{align}\label{e:protocols_inf_lambda}
u_1^{\infty}&=\phi(t)\frac{\sqrt{1+4 \pi  \left(\pi +8 \pi ^3-\sqrt{1+8 \pi ^2}\right)}}{\sqrt{2}\left(1+4 \pi ^2\right) \left(4 \pi-\sqrt{1+8 \pi ^2} \right)}\sin(\pi (\frac{1}{4}-2 y))\\
s_1^{\infty}&=\phi(t)\frac{\sqrt{1+4 \pi  \left(\pi +8 \pi ^3-\sqrt{1+8 \pi ^2}\right)}}{-2\sqrt{2}\pi \left(1 +4 \pi ^2\right)}\sin(\pi (\frac{1}{4}-2 y)).\nonumber
\end{align}  
One can see that the control strategy maximizing the performance while maintaining the cost fixed is in this case a \emph{standing wave} where chemical and mechanical drivings effectively  conspire.  Note that   the maximal performance in the presence of such cooperativity  is actually larger than in the case where only mechanical driving is present  ($P_2^{\infty}\simeq 0.44>P_2^{0}\simeq 0.08$).  The reason is that  the possibility of fast chemically driven mass redistribution (material  turnover)   complements  mechanical deformation in driving the  internal flow  which can be then  facilitated chemically even  in the presence of  large frictional forces.

\section{Chemo-mechanical crossover}\label{sec:chemo_mech}

We have seen that depending on the relative importance  of chemical vs mechanical pathways there may be  two very different optimal strategies of  actively driving the steady crawling  on a rigid substrate.  If the kinetic of turnover is much slower than that of sliding friction with the substrate,  such crawling is optimally driven by a control in the form of a \emph{traveling wave}. If, instead, mechanics is much slower than chemistry, the optimal control takes the form of a \emph{standing wave}. In this section we study the crossover between these limiting control strategies by considering the general case when the parameter  $\lambda \geq 0$ is finite. 

In our approach we will not  address the implied infinite dimensional optimal control problem in its full generality. Instead, we shall consider a finite dimensional version  of this problem by restricting the space of admissible controls to an eight-parameter space generated by an ansatz  interpolating between the traveling and the standing wave type  spatio-temporal patterns. The latter are suggested by the  solutions of the infinite dimensional limiting problems presented  in the previous sections~\ref{s:mechanical_drive}~and~\ref{s:chemical_drive}.
 
\begin {figure}[h!]
\begin {center}
\includegraphics [scale=0.45]{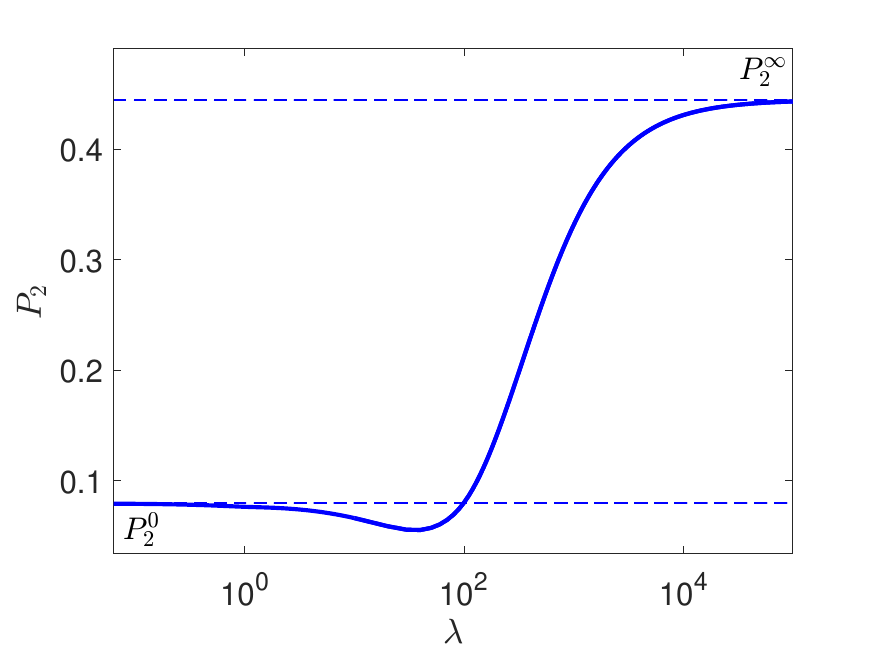}
\end {center}
\caption {Optimal performance as a function of turnover (log scale). The dashed lines indicate the analytic limits $\lambda\rightarrow 0$ and $\lambda\rightarrow \infty$ computed in Sections \ref{s:mechanical_drive}  and \ref{s:chemical_drive}.}
\label{f:optP2}
\end {figure} 

More specifically, we now consider the following parametric interpolation:
\begin{align}
s_1(y,t)=&s_{cc} \cos \left(4 \pi ^2 t\right) \cos (2 \pi  y)+s_{cs} \cos \left(4 \pi ^2 t\right) \sin (2 \pi  y)\nonumber\\ 
+&s_{sc} \sin \left(4 \pi ^2 t\right) \cos (2 \pi  y)+s_{ss} \sin \left(4 \pi ^2
   t\right) \sin (2 \pi  y),\label{e:ansatz} \\
u_1(y,t)=&u_{cc} \cos \left(4 \pi ^2 t\right) \cos (2 \pi  y)+u_{cs} \cos \left(4 \pi ^2 t\right) \sin (2 \pi  y)\nonumber\\ 
+&u_{sc} \sin \left(4 \pi ^2 t\right) \cos (2 \pi  y)+u_{ss} \sin \left(4 \pi ^2
   t\right) \sin (2 \pi  y).\nonumber
\end{align}
The goal now is to find the optimal set of  coefficients $s_{i,j}$ and $u_{i,j}$ where  the indexes   take the values either $c$ or $s$.

We use  this ansatz \eqref{e:ansatz} to analytically solve the linear problem \eqref{e:first_order}-\eqref{e:first_order_bc} and compute the function  $w_1(y,t)$. With this information at hand we can  
 directly express  the cost \eqref{e:perf1}  and also the performance \eqref{e:perf} by explicitly solving the equation \eqref{e:second_order_aver}-\eqref{e:perf_2}. We do not include the corresponding straightforward but cumbersome derivations here. The remaining Karush-Kuhn-Tucker problem of maximizing $P_2$ under the constraints $C_2=1/2$ (while respecting the inequality constraint  \eqref{e:ineq_const})  becomes finite dimensional in terms of the  height coefficients $s_{i,j}$ and $u_{i,j}$. The structure of the optimal solution can then be studied as a function of the remaining parameter $\lambda$. 
 
 In Fig.~\ref{f:optP2} we show the optimal performance $P_2(\lambda)$ computed  by numerically solving the optimization problem with an interior-point method for each value of $\lambda$. We observe a gradual  transition between the two limiting regimes. Thus, as $\lambda \to 0$ the function $P_2(\lambda)$ reaches  the value $P_2^0=1/(4\pi)$ computed analytically in  the $\lambda \ll 1$ limit. The corresponding optimal actuation strategy is the purely mechanical driving
\begin{equation} \label{L1}
s_1^0=-\pi\sqrt{2}\sin(2\pi(2\pi t-y+a))/2
\end{equation} 
 without any chemical driving:  $u_1^0=0$ found in \eqref{e:TW_actuation_no_turnover}. Note that the optimal performance  first decays to reach a minimum,  but  then finally starts to increases  reaching eventually  the plateau $P_2^{\infty}= 1/(1/\pi +12 \pi -4 \sqrt{1+8 \pi ^2})$  computed analytically in  the $\lambda \gg 1$ limit. The corresponding optimal actuation agrees with \eqref{e:protocols_inf_lambda},  but now, in view of the particular  structure of the ansatz,  with  an explicitly specified  time dependent multiplier  $\phi(t)=\sqrt{2}\sin \left(4 \pi ^2 t+a\right)$. In both limits  $a$ is an arbitrary phase.

\begin {figure}[h!]
\begin {center}
\includegraphics [scale=0.4]{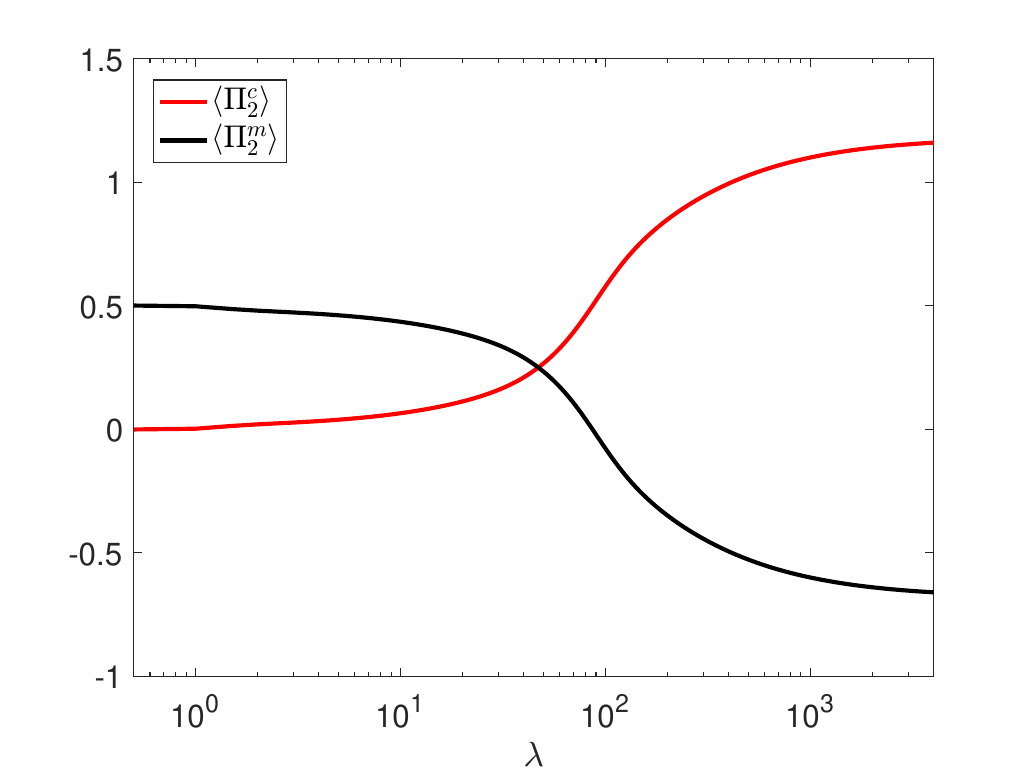}
\end {center}
\caption {Evolution of the mechanical (black) and chemical (red) costs as a function of the turnover (log scale). The sum of the two costs is maintained at the fixed value $C_2=1/2$.}
\label{f:Pi_log}
\end {figure}

In Fig.~\ref{f:Pi_log} we show the contributions of the mechanical  $\langle \Pi^m_2 \rangle$  and the chemical $\langle \Pi^c_2 \rangle$  activity to   the total energetic  cost of self-propulsion $C_2=\langle \Pi^m_2 \rangle+\langle \Pi^c_2 \rangle=1/2$. We see that in the regimes with $\lambda \ll 1$ the main  energy supply comes from the mechanical source and is represented by the work of the active stresses. Instead, in the regimes with $\lambda \gg 1$  the main   energy supply comes from the chemical ``pumps'' ensuring the appropriate target density of the turnover.  Moreover, in those regimes the flux of mechanical energy even changes sign such that mechanical actuators actually work to extract energy from the system (effectively corresponding to a brake on the global motion in the absence of chemical activity). In this way more chemical energy can be injected in. This is a consequence of the constraint fixing the total chemo-mechanical energy input and leaving the system the freedom to self-organize to optimally use this cost.

Another observable feature of the interaction between the mechanical and the chemical actuation is the  dip in the performance $P_2(\lambda) $ at  small to finite values of $\lambda$ with respect to  the performance $P_2^0$ achieved at $\lambda=0$, see Fig.~\ref{f:optP2}. This means that,  when first activated, the chemical machinery is detrimental as it interferes with the traveling wave mechanical activation and effectively works as a brake. As the corresponding  ``chemical engine'' gets sufficiently strong,   it  starts to  modify  the very regime of actuation from a traveling to a standing wave type, and the performance starts to grow reaching eventually the limit $P_2^\infty $.  

\begin {figure}[h!]
\begin {center}
\includegraphics [scale=0.45]{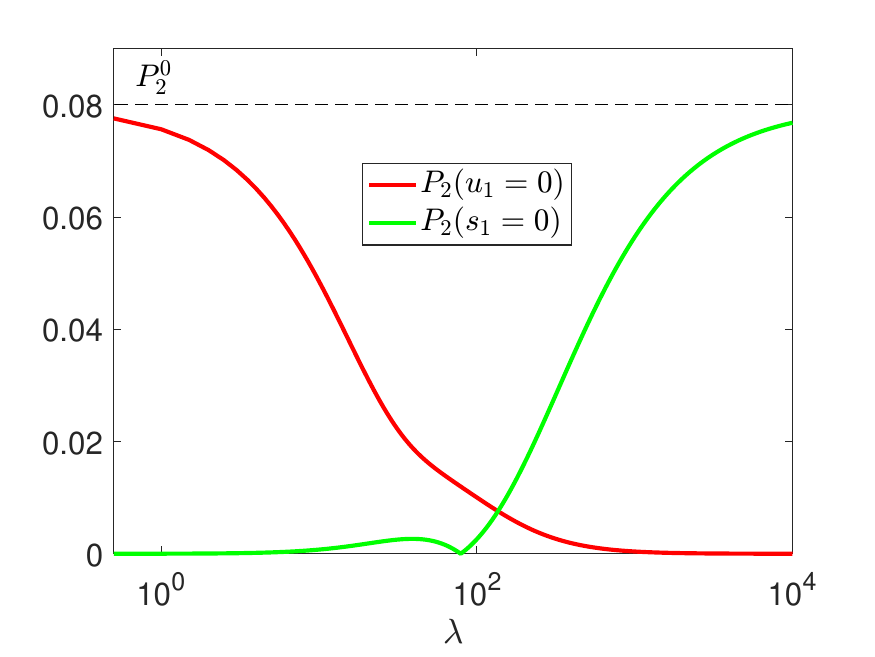}
\end {center}
\caption { Optimal performances as a function of the turnover with a mechanical driving only (red curve) and with a chemical driving only (green curve).  The cost is maintained at the fixed value $C_2=1/2$. The dashed line indicates the value of $P_2^0$.}
\label{f:mechchem}
\end {figure}

To corroborate this explanation, we show in Fig.~\ref{f:mechchem} the evolution of the optimal performance when we use a restricted actuation ansatz with either $u_1=0$ (purely mechanical actuation)  or $s_1=0$ (purely chemical actuation). As expected, the purely mechanical actuation performance only deteriorates with growing $\lambda$. Indeed, the increase of material turnover still triggers the transition from a traveling to a standing wave for the optimal actuation protocol. But the latter is associated with a decrease of performance in the absence of any active chemical recycling. Instead, the purely chemical optimal actuation, which always takes the form of a standing wave, can  take advantage of fast turnover while it is less effective than the purely mechanical actuation for slow turnover. We associate the non-monotony of the optimal performance at small to intermediate values of $\lambda$ to the somewhat arbitrary form of our  ansatz \eqref{e:ansatz} in this regime. It is nevertheless interesting that for this specific ansatz, in the limit  $\lambda\gg 1$ the purely chemical actuation  achieves exactly the same level of performance $P_2^0$ as  the purely mechanical actuation achieves at   $\lambda\ll 1$ even if  with fundamentally different spatio-temporal pattern: a traveling wave for the mechanical actuation and a standing wave  for the chemical actuation.

Another illustration of the progressive transition from a traveling wave type  to a standing wave type  actuation can be provided if we  use \eqref{e:ansatz}  to construct an ``order parameter'' type variable $\theta$, normalized to  vanish  in the standing wave regime. To this end we first rewrite 
\eqref{e:ansatz} in the form
\begin{align*}
&s_1=2s_0\cos\left( 4\pi^2t-\frac{\psi_-+\psi_+}{2}\right)\cos\left( 2\pi y-\frac{\psi_+-\psi_-}{2}\right)\\
&+( s_+-s_0)\cos\left(2\pi (2\pi t-y)-\psi_-\right) \\
&+(s_--s_0)\cos\left(2\pi (2\pi t+y)-\psi_+\right),
\end{align*}
where $s_0=\sqrt{s_{cc}^2+s_{ss}^2+s_{cs}^2+s_{sc}^2}/2$, $\Delta s=(s_{cc}s_{ss}-s_{cs}s_{sc})/2$, $ s_+=\sqrt{s_0^2+\Delta s}$, $ s_-=\sqrt{s_0^2-\Delta s}$, $\tan(\psi_+)=(s_{ss}+s_{cc})/(s_{sc}+s_{cs})$ and $\tan(\psi_-)=(s_{cc}-s_{ss})/(s_{sc}-s_{cs})$. 
This representation naturally  splits the terms  representing  a standing wave   contribution from the  two traveling waves    moving, respectively,  prograde and retrograde. Note that at  $\Delta s=0$, or equivalently 
$s_+=s_-=s_0$,
the traveling wave contribution vanishes. Therefore, the indicator 
$$\theta=\frac{( s_--s_0)^2+( s_+-s_0)^2}{4s_0^2}=1-\frac{s_++s_-}{2s_0}.$$
 compares the magnitude of the traveling wave type contributions with  the standing wave type contribution while being normalized to vanish in the purely standing wave type regime. The same type of ``order parameter'' can be constructed using  the  $u_1$ part of the of the ansatz \eqref{e:ansatz} or indeed any linear combination between $u_1$ and $s_1$.
 \begin {figure} [h!]
\begin {center}
\includegraphics [scale=0.5]{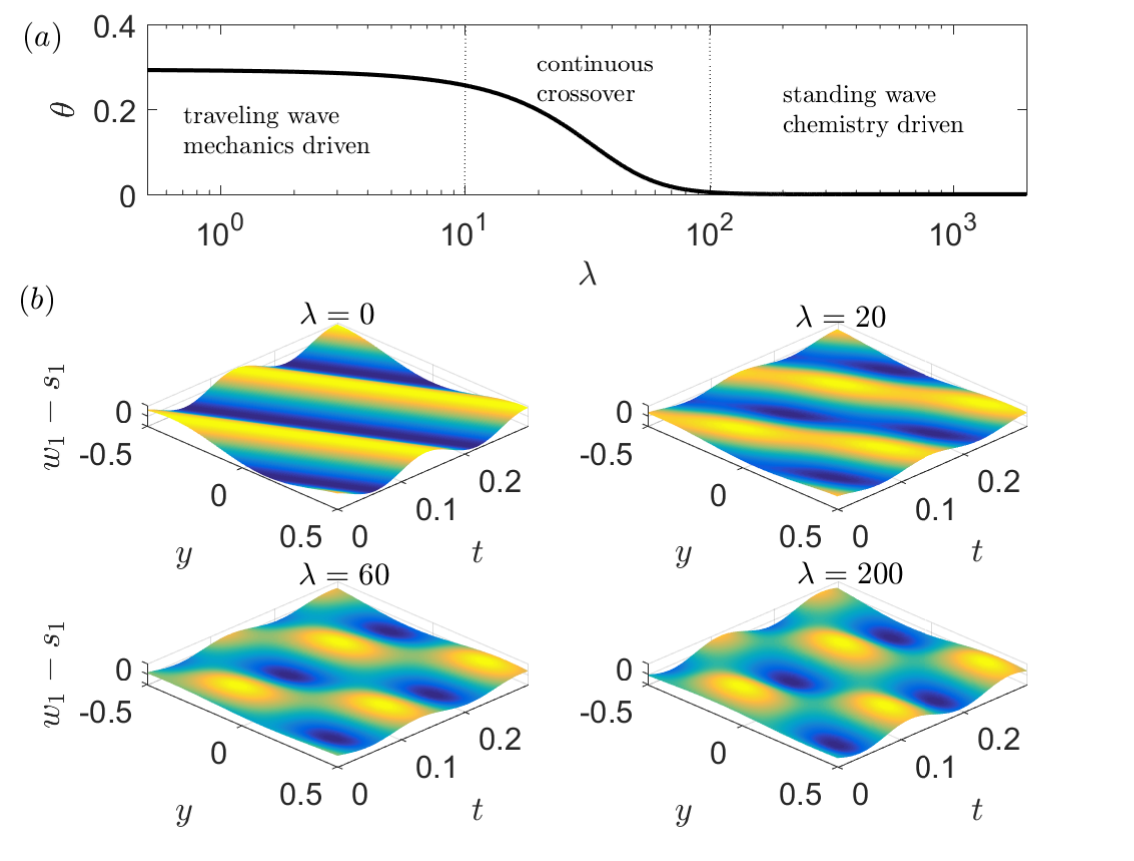}
\end {center}
\caption {Transition from a traveling wave optimal actuation to a standing wave as $\lambda$ increases. (a) Shows the decay of the order parameter $\theta$ to zero (log scale).  (b) displays the spatio-temporal density variations $w_1-s_1$ in the system at specific values of $\lambda$.}
\label{f:TWtoSWtransition}
\end {figure}  

In  Fig.~\ref{f:TWtoSWtransition} we show the behavior of the function $\theta(\lambda)$ in the regimes when both mechanical and chemical actuations are present. In the same figure we illustrate how the spatio-temporal   density profile evolves as the parameter  $\lambda$ changes from zero to infinity and the  traveling wave becomes progressively  arrested opening the way to the  formation of essentially quiescent  nodes separating periodically ``breathing'' sectors.

As we are dealing only with the simple ansatz \eqref{e:ansatz}, the exact evolution of the optimal actuation regime at a finite time scale of the mechanical friction  vis a vis the time scale of the  chemical relaxation would still have to be obtained for  general time periodic controls. For instance, it cannot be excluded that the optimal transition between the two limiting  regimes is abrupt rather than continuous, as it is suggested by our finite dimensional approximation. Also, some other  spatio-temporal patterns  may emerge along the way as one moves from one   limiting regimes to another. The rigorous clarification of all these issues which require more intense numerical approaches is left for future studies.

\section{Conclusions}\label{sec:conclusion}

We  have proposed a 1D prototypical model  of a chemo-mechanically driven system  which   can  crawl  on a solid substrate. While it can be interpreted as a paradigmatic approach to the understanding of how active stress and meshwork remodeling can conspire in living cells, it can also be used as a model of a biomimetic  self-propelling soft robot.

The  motion is actuated by two time periodic active controls which are fully intrinsic as they exert  zero average action on the system.  Those two controls are  physically fundamentally different,  with one being  mechanical and the other being chemical. The  active mechanical force field is deforming the  elastic scaffold while  the  out of equilibrium mass reservoir actively controls  the availability of building blocks of the scaffold.
 
For our analysis we have chosen a simple close-to-equilibrium  framework which enabled us to introduce the turnover kinetic time scale and the  characteristic time scale of mechanical relaxation. When the turnover kinetics is slow, the best performance at a fixed energetic cost  is achieved by a purely mechanical actuation represented by a traveling wave deformation propagating from the rear to the front of the crawler. Instead, when the turnover kinetics is fast, the best performance is reached when both mechanical and chemical drivings cooperate and form a standing wave. 
  
Our approach also allowed us to study the continuous  crossover between the  limiting  ``mechanics dominated'' and ``chemistry dominated'' regimes for a specific actuation ansatz. Here one can expect that in more elaborate models  various other intermediate regimes of actuation, characterizing alternative optimal crawling gaits, can become possible with continuous as well as discontinuous transitions between them.  The potential  complexity of this issue is already suggested by our observation  that  when the activity of the chemical reservoir is still weak, the associated  material turnover  represents itself  only as a dissipative process which  lowers the performance. However, when the chemical activity becomes sufficiently strong, the turnover enhances the performance by   offering the possibility to recycle  matter  without creating a frictional counter-flow. This simple example shows that the optimization of the metabolic actuation may  involve a complex interplay   between mechanical and chemical  active agents  and suggests that it is cooperativity of these two mechanisms that  ultimately ensures  optimality of the self-propulsion machinery. 

An  important remaining open  question is the very possibility to separate the active controls,  operating in real living systems,  into a purely mechanical actuation and a purely chemical actuation only connected to each other by the constraint that they should operate at a fixed total metabolic cost.   In crawling cells both the mechanical activity of molecular motors exerting contractile forces on the polymer network and the chemical activity regulating the turnover of the meshwork through its polymerization and depolymerization are ultimately driven by the \emph{same} chemical process: the out of equilibrium reaction of ATP hydrolysis.  The chemical and mechanical  actuations are  also tightly dynamically cross-regulated through enzyme coupled receptors \cite{Alberts2002}.  In this way  the internal  driving mechanisms are also coupled to the dynamics through a system of feedback loops.  While all these important processes  are  left outside the present  study, they should become a subject of future work.

\acknowledgments{ P.R. acknowledges the support from the french grant ANR-19-CE13-0028. L.T. acknowledges the support from the French grant ANR-10-IDEX-0001-02 PSL.}

\appendix

\bibliographystyle{apsrev4-2}%{tPHM}{elsarticle-num-names}{elsarticle-harv}
\bibliography{soft_robotics}

\end{document}